\documentclass[reprint,amsmath,amssymb,aps,prb,amssymb,superscriptaddress]{revtex4-1}
\usepackage[utf8]{inputenc}
\usepackage{color}
\usepackage{amsmath}
\usepackage{mathtools}
\usepackage{bm}
\usepackage{amsfonts}
\usepackage{dcolumn}
\usepackage{natbib}
\usepackage{bbold}
\usepackage{soul}
\usepackage[dvipsnames]{xcolor}
\usepackage{amssymb}
\usepackage[breaklinks=true,colorlinks,citecolor=blue,linkcolor=blue,urlcolor=blue]{hyperref}
\usepackage{graphicx}
\def\be{\begin{equation}}
\def\ee{\end{equation}}
\def \bea{\begin{eqnarray}}
\def \eea{\end{eqnarray}}

\usepackage{appendix}

\usepackage{mathtools}
\DeclarePairedDelimiter\bra{\langle}{\rvert}
\DeclarePairedDelimiter\ket{\lvert}{\rangle}
\DeclarePairedDelimiterX\braket[2]{\langle}{\rangle}{#1 \delimsize\vert #2}
\begin{document}

%\title{Monolayer CrTe: a room temperature ferromagnetic half-metal}
%\title{Ferroelectricity and Room Temperature Ferromagnetism in 2D Half-metallic CrTe }

\title{Room Temperature Ferroelectricity, Ferromagnetism, and Anomalous Hall Effect in Half-metallic Monolayer CrTe%; 
%/OR; \\
%\textcolor{blue}{Room temperature mutiferroicity in half-metallic CrTe monolayer}; /OR; \\
%\textcolor{brown}{CrTe Monolayer: Unlocking Room-Temperature Half-Metallic Multiferroicity}; /OR; \\
%\textcolor{purple}{Half-Metallic CrTe Monolayer: A Paradigm for Simultaneous Room-Temperature Ferroelectricity, Ferromagnetism, and Anomalous Hall Effect}
}
%%%%%% Need to change final title to SM also %%%%%%

\author{Imran Ahamed}
%\email{imranad@iitk.ac.in}
	\affiliation{Department of Electrical Engineering, Indian Institute of Technology, Kanpur 208016, India}
 \author{Atasi Chakraborty}
% \email{atasi.chakraborty@uni-mainz.de}
\affiliation{Institut f\"{u}r Physik, Johannes Gutenberg Universit\"{a}t Mainz, D-55099 Mainz, Germany}
\affiliation{Department of Physics, Indian Institute of Technology, Kanpur 208016, India}
   
\author{Pushpendra Yadav}
%\email{pyadav@iitk.ac.in}
	\affiliation{Department of Physics, Indian Institute of Technology, Kanpur 208016, India}
    
	\author{Rik Dey}
 % \email{rikdey@iitk.ac.in}
	\affiliation{Department of Electrical Engineering, Indian Institute of Technology, Kanpur 208016, India}
\author{Yogesh Singh Chauhan}
%  \email{chauhan@iitk.ac.in}
	\affiliation{Department of Electrical Engineering, Indian Institute of Technology, Kanpur 208016, India}	
\author{Somnath Bhowmick}
%\email{bsomnath@iitk.ac.in}
	\affiliation{Department of Materials Science and Engineering, Indian Institute of Technology, Kanpur 208016, India}
\author{Amit Agarwal}
\email{amitag@iitk.ac.in}
%\thanks{\\ $\dagger \dagger$ Corresponding author}	

	\affiliation{Department of Physics, Indian Institute of Technology, Kanpur 208016, India}
\date{\today}% It is always \today, today,

\begin{abstract}
Two-dimensional materials hosting ferroelectricity and ferromagnetism are crucial for low-power and high-speed information processing technologies. However, intrinsic 2D  multiferroics in the monolayer limit are rare. 
%in the monolayer limit are relatively rare. 
%Triangular lattices hold great promise for diverse magnetic phases and potential applications. Recently, room-temperature ferromagnetism was observed in triangular, few-layer CrTe crystals. 
Here, we demonstrate that monolayer CrTe, obtained by cleaving the [002] surface, is dynamically stable multiferroic at temperatures beyond room temperature. We show that it orders ferromagnetically with significant in-plane magnetocrystalline anisotropy, and it is a half-metal featuring a large half-metal gap. Remarkably, the broken inversion symmetry and buckled geometry of monolayer CrTe make it a ferroelectric with a large spontaneous out-of-plane polarization and significant magnetoelectric coupling. In addition, we demonstrate polarization or electric field-induced tunability of the anomalous Hall effect, accompanied by substantial bandstructure modulation. Our findings establish monolayer CrTe as a room-temperature multiferroic with great potential for applications in spintronics and ferroelectric devices.

\end{abstract}

\maketitle

%\tableofcontents

\section{INTRODUCTION}
\label{secI}

\begin{figure*}
\begin{center}
\includegraphics[width=\linewidth]{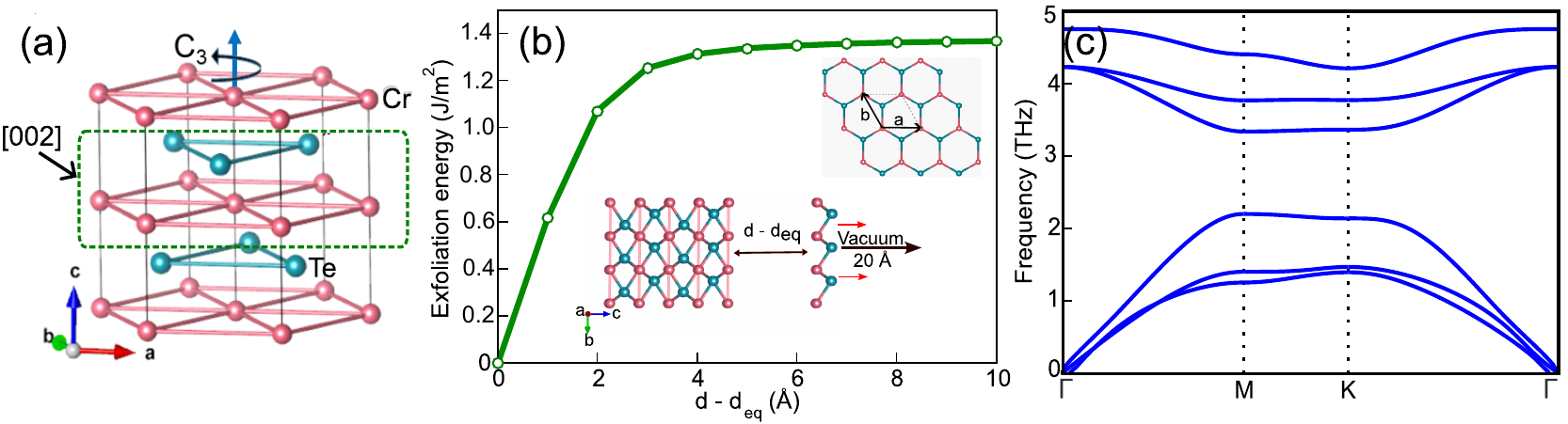}
\caption{Structure, exfoliation, and stability of monolayer CrTe. (a) NiAs-type CrTe supercell displaying the [002] monolayer within the box. (b) The exfoliation energy (1.37 J/m$^2$) of the (002)-oriented monolayer CrTe is estimated by separating one layer from the rest of the four layers. The inset shows the side and top views. (c) The phonon dispersion curve of (002)-oriented monolayer CrTe without significant negative frequencies indicates its dynamic stability.}
  \label{newfig1}
  \end{center}
\end{figure*}

Atomically thin 2D materials have attracted significant interest for diverse applications, including electronics~\cite{electronic2,electronic3,electronic4,Sinha2021B,Chakraborty2022N}, optoelectronics~\cite{opto1,opto2,Yadav_2023}, energy storage~\cite{battery},  sensors~\cite{sensor2,Hu2023T,Chakraborty2023O,Priydarshi2022S}, and catalysis~\cite{catalysis1,catalysis2,catalysis4}. However, one challenging frontier has been the quest for intrinsic 2D magnets with high Curie temperatures suitable for room-temperature magnetic applications. Several 2D materials have been proposed so far, with both the intrinsic and extrinsic origin of magnetism~\cite{PhysRevB.87.085412,doping1,doping2,doping3,doping4,proximity1,proximity2,ExfoRef,2DLimitation}. Among them, some of the 2D materials like CrI$_3$\cite{magdev3}, Cr$_2$Ge$_2$Te$_6$\cite{Cr2Ge2Te6}, VSe$_2$\cite{magdev1}, and Fe$_3$GeTe$_2$\cite{magdev2} have shown promise. However, they often fall short in terms of their transition temperatures or the strength of magnetic response.

A more significant challenge is finding multiferroic 2D materials manifesting more than one ferroic property simultaneously. One of the most desired combinations is coexisting ferroelectric (FE) and ferromagnetic (FM) ordering, having a substantial electromagnetic coupling. Although such materials hold great potential for advanced information storage and processing, offering compact form factors and low dissipation~\cite{Spaldin_2019}, they are scarce in the monolayer limit~\cite{Matsukura_2015, Spaldin_2019, Song_2022}. 2D multiferroics provide exciting possibilities for exploring novel fundamental physics via their electrically tunable structural, electronic, magnetic, and topological properties~\cite{Tokura_2014, Narayan_2018, Song_2022}. 
Additionally, they serve as versatile building blocks for creating custom-designed stacked heterostructures with tailored functionalities~\cite{Song_2022}.

Here, we predict monolayer CrTe as a remarkable intrinsic room-temperature multiferroic featuring in-plane ferromagnetism and out-of-plane ferroelectricity. Our interest in CrTe stems from the successful growth of a multi-layer CrTe via chemical vapor deposition, revealing room-temperature ferromagnetic behavior~\cite{CrTephases4}. We show that monolayer CrTe is a half-metal with a substantial gap for one spin channel while metallic for another. Our investigation into its magnetic and ferroelectric properties, within density functional theory (DFT) combined with Monte Carlo simulations, demonstrates the persistence of its robust ferromagnetism and ferroelectricity beyond room temperature~\cite{CrTephases4}. Additionally, monolayer CrTe hosts an anomalous Hall effect, indicating its topological nature. The unique out-of-plane FE polarization, which an external electric field can reverse, originates from its inversion-symmetry-breaking buckled structure. Spontaneous polarization facilitates electric field tunability of its structural, electronic, and magnetic characteristics~\cite{Buckled2D,2DHCbi, III-VFE}. These intrinsic multiferroic attributes in monolayer CrTe, combined with a significant electromagnetic coupling and anomalous Hall effect, present exciting opportunities for designing novel functionalities~\cite{mahajan2023, Spaldin_2019, Tokura_2014}.

%\section{Results}
\section{Exfoliation energy and Structural Stability}
\label{secIIA}
We first demonstrate the feasibility of exfoliating monolayer CrTe from its bulk form and assess its stability through vibrational phonon mode calculations. 
%The computational details are presented in the Supplemental Material (SM). 
{Bulk CrTe has five different phases. The crystal structure of the five phases of bulk CrTe are MnP, NaCl, NiAs, Wurtzite, and ZnS-type,~\cite{CrTephases1, CrTephases2, CrTephases3, CrTephases4}, as shown in Fig.~S1 of the Supplemental Material (SM) \footnote{\href{https://www.dropbox.com/scl/fi/m04w6gil5a0oh6pq6ac7j/SupplementaryMaterial.pdf?rlkey=tr5lqmzs66sof06f4iux42h3t&dl=0}{Supplemental Material} has details of the i) computational methods, ii) bulk magnetic phases of CrTe, iii) impact of Coulomb interactions on the ground state, and iv) the model calculation of the effective spin Hamiltonian.}. We find that the NiAs-type bulk CrTe is the most stable ground state [see Table~S1 of SM \cite{Note1}].} We construct a single-layer geometry by extracting it from the NiAs-type bulk CrTe along the [002] direction [see Fig.~\ref{newfig1}\textcolor{blue}{(a)}]. For ease of notation, in the rest of the manuscript, we refer to the (002)-oriented CrTe monolayer as monolayer CrTe. To establish the feasibility of exfoliation of monolayer CrTe, we calculate its exfoliation energy ($E_{ex}$). For this, we consider a five-layered CrTe system in equilibrium and separate one layer by increasing its distance from the rest of the layers, as shown in the inset of Fig.~\ref{newfig1}\textcolor{blue}{(b)}. We estimate the exfoliation energy using the expression \cite{ExfoEn}, 
\begin{equation}
    E_{ex} = \frac{E^{d-d_{eq}}_{n} - E^{d_{eq}}_{n}}{A}~.\label{eq:1}
\end{equation}
Here, $A$, $d_{eq}$ and $E^{d_{eq}}_{n}$ represents the in-plane surface area, equilibrium layer separation in a bulk crystal, and the total energy of the five-layer slab in equilibrium, respectively. 
%For our first principle calculations, we choose a five layer CrTe slab. 
In Eq.~\eqref{eq:1}, $E^{d-d_{eq}}_{n}$ is the total energy of the system when one monolayer of the five-layer thick slab is at $d-d_{eq}$ distance from the rest. The details of the first principles-based density functional theory calculations are presented in SM \cite{Note1}. We estimate the exfoliation energy of monolayer CrTe to be 1.37 J/m$^2$ [see Fig.~\ref{newfig1}\textcolor{blue}{(b)}].  This is comparable in magnitude to the exfoliation energy of GeP$_3$ and InP$_3$~\cite{GeP3, InP3}. The non-vdW nature of bulk CrTe~\cite{ExfoRef} leads to the larger exfoliation energy of CrTe monolayer compared to other known vdW materials such as graphene, MoS$_2$, phosphorene, SnP$_3$, GdTe$_3$, GeS, GeTe and Cu(Cl, Br)Se$_2$ \cite{ExfoEn, GdTe3}.   

\begin{figure*}[t]
\begin{center}  \includegraphics[width=0.8\linewidth]{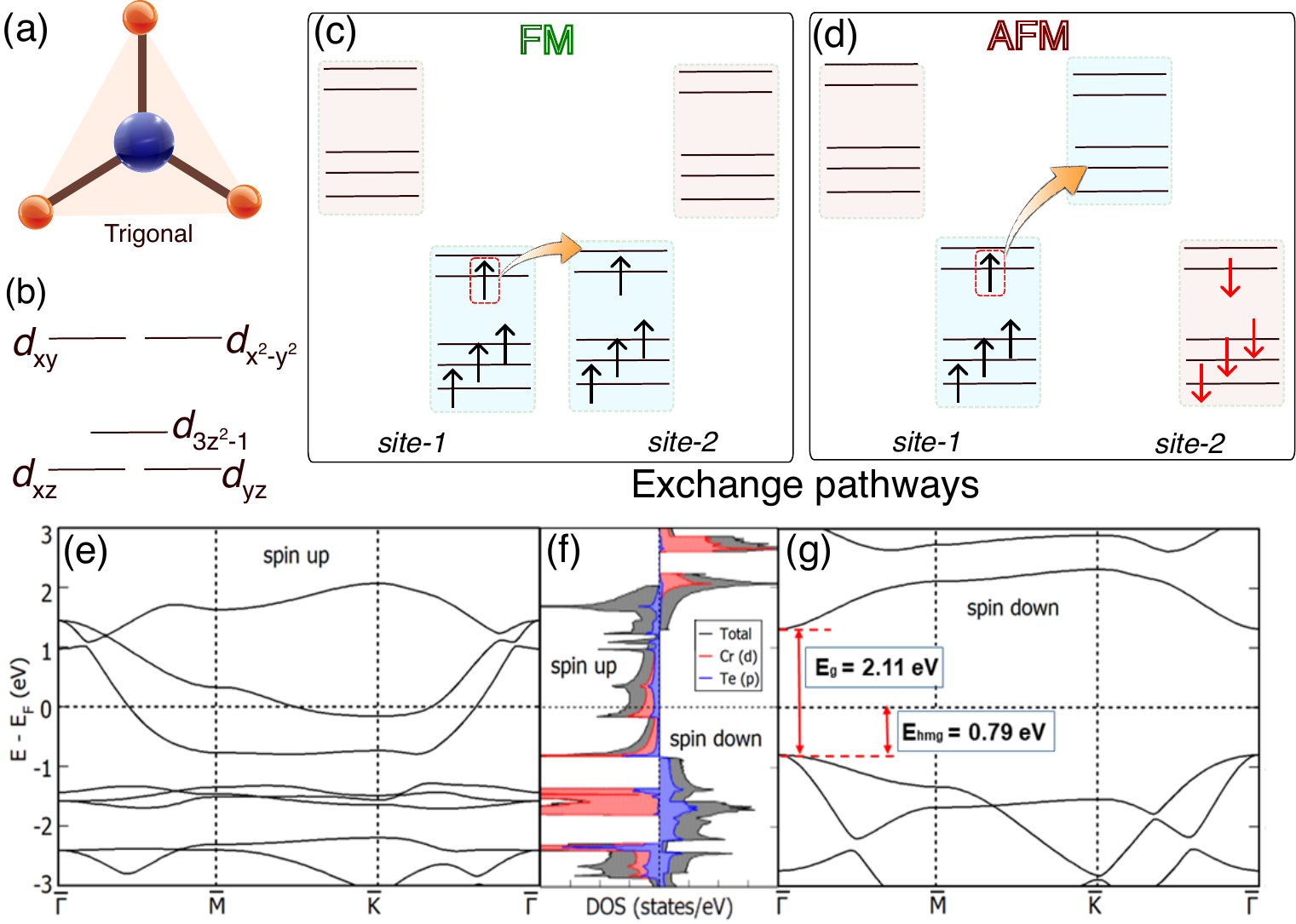}
\caption{(a) The trigonal environment of Cr formed with nearest neighbor Te atoms. (b) The crystal field splitting of $d$ orbitals within an ideal trigonal environment. Schematic representation of (c) ferromagnetic and (d) antiferromagnetic exchange mechanisms between two neighboring Cr sites. The arrow indicates the hopping paths. (e-g) The energy dispersion indicates half-metallic nature and density of states (DOS). The states near the Fermi energy have large contributions from Cr-$d$ states.
\label{newfig2}}
\end{center}
\end{figure*}

To ascertain the dynamical stability of the free-standing monolayer, we check that its phonon dispersion has no imaginary component. For this, we perform the phonon dispersion calculation through the linear response method within density functional perturbation theory (DFPT) \cite{DFPT} as implemented in the PHONOPY code \cite{phonopy}. We show the calculated phonon dispersion of monolayer CrTe in Fig.~\ref{newfig1}\textcolor{blue}{(c)}. The monolayer has the exact stoichiometry as bulk CrTe, containing one Cr and one Te atom per unit cell. Thus, the phonon dispersion of monolayer CrTe has six modes with three acoustic and three optical branches, as highlighted in Fig.~\ref{newfig1}\textcolor{blue}{(c)}. We find that monolayer CrTe has positive phonon frequencies throughout the Brillouin zone, indicating that it is dynamically stable. The possibility of exfoliating monolayer CrTe in a stable form and its dynamical stability has not been demonstrated earlier.

\section{Magnetism and Ground State Anisotropy in monolayer $\textrm{CrTe}$}
\label{secIIB}
The Cr atoms form a local trigonal network with the neighboring Te within the monolayer geometry, as depicted schematically in Fig.~\ref{newfig2}\textcolor{blue}{(a)}. The crystal field splitting of $d$ orbitals within an ideal trigonal network is shown in Fig.~\ref{newfig2}\textcolor{blue}{(b)}. However, Cr atoms are not in the plane of Te atoms, leading to a breaking of degeneracy in the Jahn-Teller active orbitals of the ideal trigonal network. This can allow for both ferromagnetic and antiferromagnetic (AFM) ordering. Schematic representations of the hopping mechanisms for both FM and AFM ground states are depicted in Fig.~\ref{newfig2}\textcolor{blue}{(c)} and \textcolor{blue}{(d)}, respectively.  %showing ferromagnetic (FM) state to be the ground state.  

\begin{table}[b]
\centering
\caption{\label{Table1} The relative energies of different spin configurations such as FM, AFM-1, and AFM-2 with spins aligned along $x$, $y$, and $z$ direction, with respect to the ground state spin configuration. In the last column, we report the exchange interaction parameter $J$ along different axes, considering a nearest neighbor Heisenberg model.\\}

\begin{tabular}{c c c c c}
\hline \hline 
  {\begin{tabular}[c]{@{}c@{}}Spin \\direction\end{tabular}} &
  {\begin{tabular}[c]{@{}c@{}}E$_{FM}$ \\(meV)\end{tabular}} &
  {\begin{tabular}[c]{@{}c@{}}E$_{AFM-1}$ \\(meV)\end{tabular}} &
  {\begin{tabular}[c]{@{}c@{}}E$_{AFM-2}$ \\(meV)\end{tabular}} &
  {\begin{tabular}[c]{@{}c@{}}J \\(meV)\end{tabular}} \\
  \hline
 \begin{tabular}[c]{@{}c@{}} $x$\\ $y$\\ $z$\end{tabular} & 
  \begin{tabular}[c]{@{}c@{}} 0.00\\ 0.04\\ 13.29\end{tabular} &
  \begin{tabular}[c]{@{}c@{}} 781.82\\ 779.33\\ 776.72\end{tabular} &
  \begin{tabular}[c]{@{}c@{}} 777.67\\ 782.97\\ 776.65\end{tabular} &
  \begin{tabular}[c]{@{}c@{}} 8.14\\ 8.12\\ 7.95\end{tabular} \\
  \hline \hline
\end{tabular}
\end{table}

To identify the magnetic ground state of the system, we perform self-consistent total energy calculations for different magnetic configurations. Our density functional theory (DFT) based calculations reveal that the ferromagnetic (FM) state is lower in energy than the antiferromagnetic (AFM) states with an energy difference of $\sim$780 meV [see Table~\ref{Table1}]. We consider FM and two different AFM spin configurations in a $2 \times 2$ supercell to calculate the exchange parameters [see Fig.~S3 in SM \cite{Note1}]. The calculated total moment per Cr atom is 4 $\mu_B$, consistent with a high spin configuration as shown in Fig.~\ref{newfig2} \textcolor{blue}{(c)} and \textcolor{blue}{(d)}. 

The electronic band structure and density of states (DOS), calculated using the generalized gradient approximation (GGA) with the Hubbard $U$ correction, or the GGA+\textit{U} scheme, with $U=2$ eV, show that the FM state of monolayer CrTe is half metallic [see Fig.~\ref{newfig2}\textcolor{blue}{(e)} and \textcolor{blue}{(g)}]. Half metallicity is also found in NiAs-type bulk CrTe (see Fig.~S2 in SM). The up-spin channel has a finite contribution at the Fermi level, as seen in Fig.~\ref{newfig2}\textcolor{blue}{(e)}. However, the down-spin channel is insulating with a large direct band gap of 2.11 eV; see Fig.~\ref{newfig2}\textcolor{blue}{(g)}. We find the states near the Fermi energy are dominated by contributions from the Cr-$d$ orbital along with negligible hybridization from the Te-$p$ states [see Fig.~\ref{newfig2}\textcolor{blue}{(f)}]. We estimate the half-metal energy gap for monolayer CrTe to be $E_{\rm hmg}\sim$ 0.79 eV, which is three times higher than that of the bulk ($E_{\rm hmg}\sim$ 0.24 eV). The $E_{\rm hmg}$ of 0.79 eV for CrTe-monolayer is comparable to that of 1T-TaN$_2$ monolayer ($E_{\rm hmg}\sim$ 0.72 eV)~\cite{TaN2}. This 100\% spin polarization at the Fermi energy with a large half-metallic gap in monolayer CrTe makes it an exceptional candidate for potential applications in spin-filter and spin transport devices, spin-selective optical excitations and opto-magnonic devices.

\begin{figure*}[t]
\begin{center}
\includegraphics[width=\linewidth]{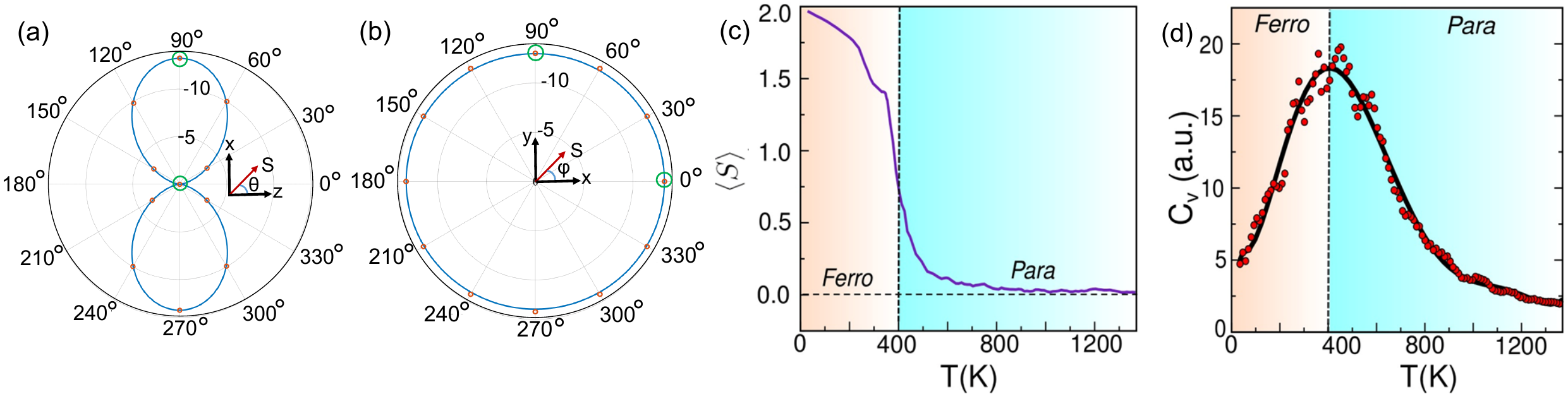}
\caption{Magnetocrystalline anisotropy plots in (a) $xz$ and (b) $xy$ plane, calculated using DFT (red data points), and showing in-plane orientation to be energetically favorable. The data points enclosed in green circles are used to determine the anisotropic exchange parameters. Classical Monte Carlo simulation predicts a magnetic phase transition temperature $T_c$ around 400 K based on (c) magnetic moment vs. temperature and (d) specific heat ($C_v$) vs temperature plots. The solid black line is the best-fit curve, serving as a visual aid.\label{newfig3}}
\end{center}
\end{figure*}  

Since long-range magnetic ordering in low dimensional systems is supported by the magnetocrystalline anisotropy, we have calculated magnetocrystalline anisotropy energy (MAE) for the ground state FM state. The MAE ($\theta,\phi$) is defined as the change in energy when spin orientation [$\hat{S}(\theta, \phi)$, $\theta$ and $\phi$ being polar and azimuthal angle] of the system is varied from its ground state direction (easy axis). The MAE for monolayer CrTe is presented in Fig.~\ref{newfig3}\textcolor{blue}{(a)} and \textcolor{blue}{(b)}. The energy is lowest when the out-of-plane $z$-component of the spin vanishes, and monolayer CrTe possesses an easy plane anisotropy with MAE(90$^o$,0$^o$) = -3.3~meV per unit cell [see Fig.~\ref{newfig3}\textcolor{blue}{(a)}]. The anisotropy in the $xy$ plane is negligible [see Fig.~\ref{newfig3}\textcolor{blue}{(b)}]. To put this in perspective, we note that the MAE  of CrTe monolayer is comparable to or higher than that of several other well studied 2D ferromagnetic monolayers, such as Hf$_2$Te (2.2 meV/u.c) \cite{Hf2Te}, Fe$_3$GeTe$_2$ (2.76 meV/u.c) \cite{Fe3GeTe2}, CrI$_3$ (1.6 meV/u.c) \cite{CrI3}, Fe$_3$P (2.45 meV/u.c) \cite{Fe3P}, Fe$_2$Si (1.15 meV/u.c) \cite{Fe2Si}, MnAs (0.56 meV/u.c) \cite{MnAs}, CrPbTe$_3$ (1.37 meV/u.c) \cite{CrPbTe3}.

\section{Paramagnetic to Ferromagnetic phase Transition}
\label{secIIC} 
To estimate the transition temperature of the FM phase in CrTe,  
we map the system to the nearest neighbor ($nn$) Heisenberg model using the ground state energies obtained for different magnetic configurations [see SM \cite{Note1} section-IV for details]. Since the second neighbor's distance (7.21 \AA) is substantially larger than the nearest neighbor's (4.16 \AA), we restrict our spin model to the first neighbor. Furthermore, we find that the calculated single ion anisotropy term [$\propto (S_i^\alpha)^2$, with $\alpha = x,~y,~z$] for the CrTe monolayer is minimal and can be safely neglected [see SM section-IV for details]. We obtain the effective spin model to be,  
\begin{equation}\label{eq:2}
H = -\frac{1}{2}\sum_{i\neq j} [J^x_{nn}S^x_iS^x_j + J^y_{nn}S^y_iS^y_j + J^z_{nn}S^z_iS^z_j]~. %+ \sum_i A_i(S^m_i)^2 
\end{equation}
Here, $J_{nn}^\alpha$ (where $\alpha=x,y,z$)  represent the $nn$ exchange interactions along the $\alpha$- axis. The values of these exchange parameters along the three axes are tabulated in the last column of Table~\ref{Table1}. As a consistency check, we compare the MAE$(\theta,\phi)$ values obtained using the exchange interaction parameters from Eq.~\eqref{eq:2} with the DFT calculated MAE. We find that MAE values obtained from the model Hamiltonian [solid blue curve in Fig.~\ref{newfig3}\textcolor{blue}{(a)} and \textcolor{blue}{(b)}] agree well with the DFT calculated values [red data points]. %, which further confirms that the Heisenberg spin Hamiltonian describes the monolayer system effectively. 
Motivated by this, we use the obtained magnetic model to determine the Curie temperature.

\begin{figure*}
\begin{center}
\includegraphics[width=1\linewidth]{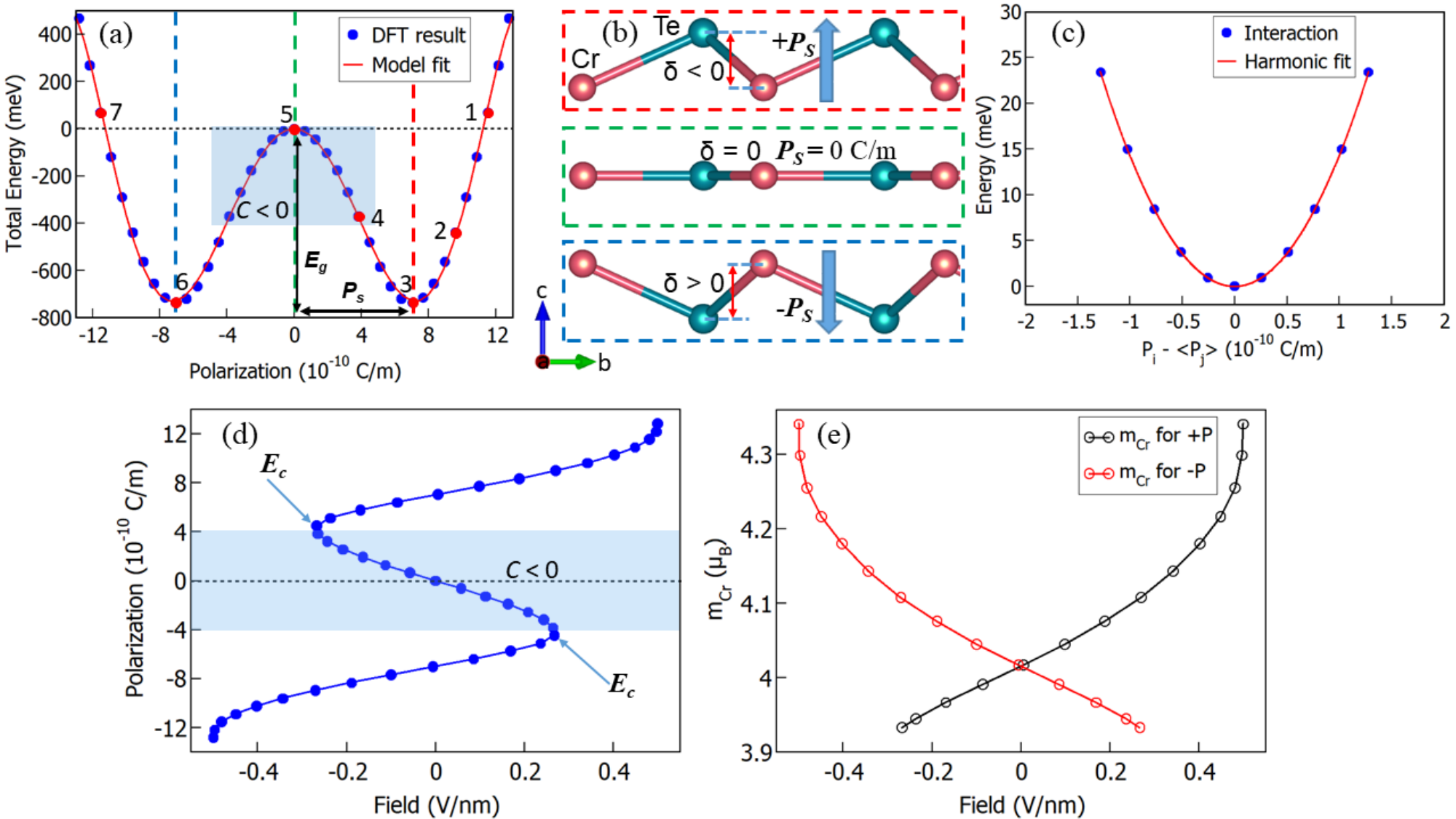}
\caption{(a) The polarization dependence of the relative total energy showing the spontaneous polarization \textit{$P_s$} and polarization switching barrier height \textit{$E_G$}. (b) The buckled CrTe honeycomb monolayer's side view shows polarization's dependence on buckling height $\delta$. (c) The nearest neighbor dipole-dipole interaction energy of monolayer CrTe is calculated by mean-filed theory showing the harmonic fit. (d) The polarization-electric field curve (S-curve) of the FE monolayer is obtained from the data of (a). (e) The variation of the magnetic moment of Cr-atom with the electric field.}
  \label{Fig4}
  \end{center}
\end{figure*}

To estimate the magnetic transition temperature ($T_c$), we do classical Monte Carlo (MC) simulations with the nearest neighbor spin Hamiltonian specified by Eq.~\eqref{eq:2}. We start with randomly oriented Heisenberg spins on a system of 100$\times$100 spins, allowing all spins to rotate freely in three-dimensional space. We choose $10^7$ relaxation steps to reach equilibrium at each temperature, then calculate the average energy value and moment over the next $10^5$ steps. We present the total spin per Cr site, $\langle S \rangle$, as a function of temperature in Fig.~\ref{newfig3}\textcolor{blue}{(c)}. For calculating the temperature dependence of the specific heat, we used the following expression,
\begin{equation}\label{eq:3}
    C_v \approx \frac{\langle E^2 \rangle - \langle E \rangle ^2}{T^2}~.
\end{equation}
We display the variation of the calculated $C_v$ per site with temperature in Fig.~\ref{newfig3}\textcolor{blue}{(d)}. Both spin and specific heat plots show a clear signature of paramagnetic to ferromagnetic transition, with $T_c \sim400~K$  [Fig.~\ref{newfig3}\textcolor{blue}{(c)} and \textcolor{blue}{(d)}]. Earlier experimental work reported the $T_c$ of thin films of CrTe crystals of thickness 100 nm, 69 nm, and 37 nm crystals to be $\sim$340, $\sim$370, and $\sim$360 K, respectively~\cite{ExfoRef}. Our calculations and these results strongly suggest that layered samples of CrTe are room-temperature ferromagnets down to the monolayer limit. 

Having shown the stability and the possibility of room temperature ferromagnetism in monolayer CrTe, we now focus on demonstrating the existence of ferroelectricity in monolayer CrTe in the next section. 

\begin{figure*}[ht]
\begin{center}
\includegraphics[width=\linewidth]{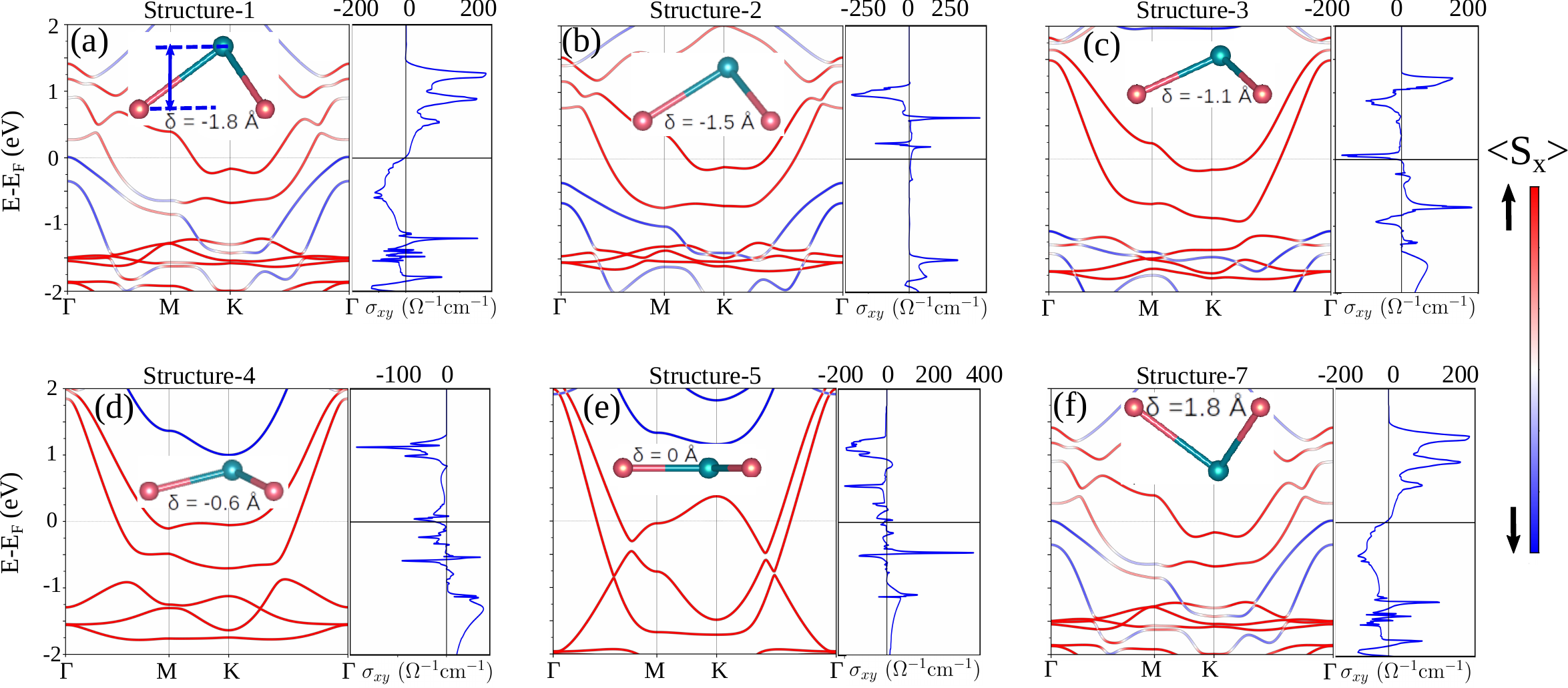}
\caption{Impact of applied electric field on the electronic band structure and the spin polarization (left) and anomalous Hall conductivity $\sigma_{xy}$ (right) for different atomic arrangements. The color scale in the bandstructure plots indicates the $s_x$ polarization of the itinerant spins. Panel (a) is the equilibrium structure with $\delta=1.8$ \AA\, which gradually decreases to $\delta=0$ (zero polarization state), as shown in panels (b) - (e), resulting in a significantly reduced AHC. Panel (f) is the same as (a) but buckled in the opposite direction.}
\label{newfig5}
\end{center}
\end{figure*}

\section{Out-of-Plane Ferroelectricity in monolayer $\textrm{CrTe}$}
\label{secIID}
2D materials with no inversion symmetry can have an out-of-plane polar axis and show electric polarization leading to ferroelectricity~\cite{Buckled2D,III-VFE}.  
%as shown in the InP series of materials~\cite{III-VFE}, \textcolor{red}{among others~\cite{xx}}. 
The buckled nature of the monolayer CrTe, highlighted in the inset of Fig.~\ref{newfig1}\textcolor{blue}{(b)}, motivates the exploration of its FE properties. Using the optimized buckled monolayer of CrTe, we study the FE properties utilizing the first-principle Berry phase method~\cite{King-smith1993, Berry1984, King-smith1994, Resta1994}. The change in buckling height between Cr and Te planes (denoted by $\delta$) changes the electric dipole, resulting in the variation of the FE polarization. Figure~\ref{Fig4}\textcolor{blue}{(a)} illustrates the variation of the relative total energy of the monolayer CrTe with the change in the out-of-plane FE polarization. The variation of the total energy with polarization has the form of an anharmonic double well potential (w-curve) with two minima corresponding to the top and bottom panels of Fig.~\ref{Fig4}\textcolor{blue}{(b)}. The maximum energy point in the w-curve corresponds to the planar monolayer structure shown in the middle panel of Fig.~\ref{Fig4}\textcolor{blue}{(b)}. The two minima points correspond to the negative and positive out-of-plane FE polarization, achieved by the inversion of the buckling.

We quantify the FE behavior of monolayer CrTe using the Landau-Ginzburg (LG) equation. The LG equation for the free energy (\textit{G}) as a function of polarization (\textit{P$_i$}) per unit cell is~\cite{LG1,LG2,LG3}
\begin{equation}\label{eq:4}
G = \sum_i\Bigg[\frac{A}{2}P_i^2 + \frac{B}{4}P_i^4 + \frac{C}{6}P_i^6\Bigg] + \frac{D}{2}\sum_{\langle i,j\rangle}(P_i - P_j)^2 ~.
\end{equation}
Here, \textit{i} indicates the unit cells, and \textit{j} indicates the nearest neighbor unit cell of the \textit{i}$^{th}$ unit cell. The free energy is in units of meV per unit cell, and \textit{P} is in units of $10^{-10}$ C/m. We estimate the coefficients \textit{A}, \textit{B}, and \textit{C} by fitting the DFT data with the Eq.~\eqref{eq:4}. We obtain the  values of the coefficients \textit{A}, \textit{B}, and \textit{C} to be $-61.71$ meV/($10^{-10}$ C/m)$^2$, 1.51 meV/($10^{-10}$ C/m)$^4$  and $-0.006$ meV/($10^{-10}$ C/m)$^6$, respectively. The spontaneous polarization and potential barrier are \textit{$P_s$} = 7 $\times$ $10^{-10}$ C/m and \textit{$E_G$} = 729 meV, respectively. The value of \textit{$P_s$} for monolayer CrTe is higher than that of some known 2D FEs such as \textit{MX} (\textit{M} = Ge, Sn; \textit{X} = S, Se) \cite{LG1}, $\delta$-(As,Sb,Bi)N \cite{AsN}, group III–V monolayers (III = Ga, In; V = P, As, Sb)  \cite{III-VFE},  other 2D honeycomb binary compounds \cite{2DHCbi} and some elemental group V monolayers \cite{EleV2D}. The energy barrier for polarization switching is higher than most of the known 2D FE materials and comparable to the value in two-dimensional In$_2$Se$_3$ \cite{In2Se3FE}. 

The $S$-curve, illustrated in Fig.~\ref{Fig4}\textcolor{blue}{(d)}, captures the polarization dependence on the electric field, and it is determined by $E = dG/dP$~\cite{Hoffmann2019, Priydarshi2022S}. The blue shaded region in Fig.~\ref{Fig4}\textcolor{blue}{(a)} and Fig.~\ref{Fig4}\textcolor{blue}{(d)} is the region of negative capacitance. The capacitance $C$ is proportional to the slope $dP/dE$. This region of negative capacitance in the $S$-curve is difficult to observe experimentally in ferroelectrics because of the unstable nature of the material around $P = 0$ (i.e., paraelectric region). The $S$-curve gives the coercive field $E_c$ of $\pm$ 0.27 V/nm corresponding to the field required to switch the polarization direction.

The FE transition temperature is dictated by the average dipole-dipole interaction strength between the nearest neighboring unit cells or the parameter \textit{D} in Eq.~\eqref{eq:4}. We calculate \textit{D} by taking a supercell of the dipole and fitting the relative total energy with a classical harmonic oscillator-like parabolic function \textit{D}~\cite{LG1, MeanfieldD} as indicated in Eq.~\eqref{eq:4} and shown in Fig.~\ref{Fig4}\textcolor{blue}{(c)}. 
%having coupling strength \textit{D}~\cite{LG1,MeanfieldD}. 
We obtain \textit{D} to be 7.13 meV/($10^{-10}$ C/m)$^2$. %Following the method of Fei \textit{et} \textit{al}. \cite{LG1}, 
We can estimate the transition temperature of the FE from the expression \cite{LG1} $T_c \sim DP_s^2/k_B$. We estimate the $T_c$ for monolayer CrTe to be much higher than room temperature (around 4000 K). This indicates that monolayer CrTe is a room-temperature multiferroic, supporting an out-of-plane FE order and an in-plane ferromagnetic half-metallic state. %polarization, 
%The high value of coupling strength \textit{D} will counter the thermal fluctuations in the FE structure, thus giving rise to a large transition temperature \cite{EleV2D}. 

The simultaneous presence of ferroelectricity and ferromagnetism in monolayer-CrTe motivates the study of the magnetoelectric (ME) coupling and the anomalous Hall effect. A finite ME coupling 
indicates the possibility of tuning i) the magnetic state by applying an electric field and ii) the polarization state by applying a magnetic field. More interestingly, time reversal symmetry broken magnetic systems can also host anomalous Hall effect (AHE). The AHE can be tuned in such a scenario by applying a vertical electric field. We explore both these effects in the following two sections. 

\section{magnetoelectric coupling in monolayer $\textrm{CrTe}$}

To understand the ME coupling, we analyze the electric field data, polarization value, and the magnetic moment of the Cr-atom corresponding to the respective polarization state. In the stable region of the FE, the variation of the magnetic moment of the Cr-atom in the positive as well as the negative polarization state is shown in Fig.~\ref{Fig4}\textcolor{blue}{(e)}. We find that the ME coupling is positive for the up-polarization state and negative for the down-polarization state based on the slope of the curve. The ME coupling coefficient $\alpha$ is proportional to the variation of the magnetic moment ($\Delta m$) with the change in the applied electric field ($\Delta E$), or  $\alpha = \frac{\mu_0\Delta m}{\Delta E}~$. Here, $\mu_0$ is the permeability of free space. The linear ME coupling coefficient is calculated for the linear region of Fig.~\ref{Fig4}\textcolor{blue}{(e)} for an electric field ranging from $-0.3$ to $+0.3$ V/nm. We obtain the value of $\alpha$ to be $+9.4$ ps/m and $-9.4$ ps/m for the positive and negative $P_z$ state, respectively. This significant value of the ME coupling $\alpha$ is of the same order of magnitude as that found in bulk CaMnO$_3$ and in the prototypical ME material -  bulk Cr$_2$O$_3$~\cite{CaMnO3ME, Cr2O3ME}.

\section{Electrically Tunable Anomalous Hall Conductivity}
\label{secIIE}
The anomalous Hall conductivity in 2D time reversal symmetry broken materials is the Hall response without a magnetic field~\cite{AHC_Karplus1954}. It is given by the Brillouin zone (BZ) integral of the out-of-plane Berry curvature of a given band ($\Omega_{n}^z$) weighted by its  occupation factor~\cite{Chang1995,Berry_Xiao2010},
\bea \label{eq:6}
\sigma_{xy}^{\rm AHC} = -\frac{e^2}{\hbar} \sum_{n} \int_{\rm BZ} \frac{d\textbf{k}}{(2\pi)^2}f_{n}(\textbf{k})\Omega_{n,z} (\textbf{k})~.
\eea
Here, $f_{n}(\textbf{k}) = 1/[1+e^{(E_n(\textbf{k}) - \mu)/(k_BT)}]$ is the equilibrium Fermi function of the system with chemical potential $\mu$ and temperature $T$. The Berry curvature of the $n^{th}$ band in a system can be calculated using~\cite{Berry1984},
\begin{equation}\label{eq:7}
    \Omega_{n}^z =-\mathrm{2Im}\sum_{m\ne n}\frac{\bra{\psi_{m\textbf{k}}}\partial_{k_x}H\ket{\psi_{n\textbf{k}}}\bra{\psi_{n\textbf{k}}}\partial_{k_x}H\ket{\psi_{m\textbf{k}}}}{(E_n-E_m)^2}~.
\end{equation}
Here, $E_n$ and $E_m$ are the energy eigenvalues of the $n^{th}$ and $m^{th}$ band, respectively. 

We highlight the vertical electric field-induced modulation of the electronic band structure, spin polarization of the itinerant electrons, and the AHC in Fig.~\ref{newfig5}. The vertical electric field controls the polarization state or the buckling of the monolayer. This is reflected in the modulation of electronic properties, spin polarization, and the AHC. The electric field-induced modulation of the magnetic moments of the ferromagnets is shown in Fig.~\ref{Fig4}\textcolor{blue}{(e)}. This makes monolayer CrTe a versatile platform with electrically tunable physical properties, including the AHC. This can be very useful for electronic, spintronic, and optoelectronic devices.

\section{Conclusion} \label{secIII}

In summary, we demonstrated monolayer CrTe to be a dynamically stable room temperature multiferroic at the atomic scale. It exhibits fascinating properties, featuring both half-metallicity with 100\% spin polarization induced by in-plane ferromagnetism and strong out-of-plane ferroelectricity. These properties make monolayer CrTe a promising candidate for various applications. 
Its half-metallic nature makes monolayer CrTe well-suited for spintronic applications like spin injection and spin filtering. The presence of ferroelectricity makes it an excellent choice for FE memory devices and tunnel field-effect transistors.

More interestingly, we establish the existence of a 
robust ME coupling along with
the presence of an anomalous Hall state in monolayer CrTe. The coexistence of anomalous Hall and half-metallicity indicates 
%The discovery of a robust magnetoelectric coupling, along with the observation of an anomalous Hall state, suggests 
the presence of topologically protected edge states that are 100\% spin polarized. %, in nano-ribbon and other confined geometries of monolayer CrTe. 
These edge states can carry dissipationless charge and spin currents in nano-ribbon and other confined geometries of monolayer CrTe.  The multiferroic characteristics and strong ME coupling of CrTe monolayers offer exciting prospects for magnetic sensors, information devices, and high-density data storage applications. Additionally, CrTe provides a unique atomic monolayer platform with electric field tunable electronic, magnetic, anomalous Hall, and ferromagnetic properties. This opens up several exciting directions for further studies.

\begin{acknowledgments}
IA acknowledges IIT Kanpur for providing the Post-doctoral Fellowship. AC acknowledges the Science and Engineering Research Board (SERB), India, for National Postdoctoral Fellowship (PDF/2021/000346) and the Alexander von Humboldt Foundation, Germany, for partial financial support. Authors acknowledge financial support from SERB (Grant No. CRG/2021/003687, MTR/2019/001520, CRG/2018/002400), Swarnajayanti Fellowship (Grant No. DST/SJF/ETA-02/2017-18), Department of Science and Technology, India (DST/NM/TUE/QM-6/2019(G)-IIT Kanpur). Authors acknowledge the National Supercomputing Mission (NSM) for providing computing resources of “PARAM Sanganak” at IIT Kanpur, which is implemented by C-DAC and supported by the Ministry of Electronics and Information Technology (MeitY) and Department of Science and Technology (DST), Government of India. The authors also acknowledge the HPC facility provided by CC, IIT Kanpur.
\end{acknowledgments}

\bibliography{bibliography}

\end{document}